\documentclass[12pt,twoside]{article}
\usepackage{fleqn,epsfig,espcrc1}
\usepackage{graphicx}

\def\endauthors{}
\def\authors#1\endauthors{#1}

\def\be{\begin{equation}}
\def\ee{\end{equation}}
\def\br{\begin{eqnarray}}
\def\er{\end{eqnarray}}
\def\brn{\begin{eqnarray*}}
\def\ern{\end{eqnarray*}}
\def\rf#1{{(\ref{#1})}}
\def\ket#1{|#1 \rangle}
\def\bra#1{\langle #1|}

\def\ie{{\em i.e., }}

\def\lbar{\mbox{$\lambda$\kern-0,450em \vrule width0,35em height1,252ex
depth-1,21ex \kern0,051em}}
\def\dbar{\mbox{d\kern-0,347em \vrule width0,3em height1,252ex depth-1,21ex
\kern0,051em}}
\def\Dbar{\mbox{D\kern-0,735em \vrule width0,3em height0,86ex depth-0,81ex
\kern0,40em}}

\def\lsim{\:\raisebox{-0.5ex}{$\stackrel{\textstyle<}{\sim}$}\:}

\def\pb {{\bf p}}
\def\qb {{\bf q}}

\def\Qb{ {\bf Q}}

\def\d{\dagger}

\def\b {{\beta}}
\def\e {{\epsilon}}
\def\g {{\gamma}}

\def\m {{\mu}}
\def\n {{\nu}}

\def\w {{\omega}}

\def\C {{{\cal C}}}
\def\D {{{\cal D}}}

\def\I {{{\cal I}}}

\def\ba#1{\begin{array}{#1}}
\def\ea{\end{array}}

\def\be{\begin{equation}}
\def\ee{\end{equation}}
\def\br{\begin{eqnarray}}
\def\er{\end{eqnarray}}
\def\brn{\begin{eqnarray*}}
\def\ern{\end{eqnarray*}}
\def\bit{\begin{itemize}}
\def\eit{\end{itemize}}
\def\bnu{\begin{enumerate}}
\def\enu{\end{enumerate}}
\def\x{\times}
\def\={{\simeq}}

\def\fot{\frac{1}{2}}

\def\go{\rightarrow  }

\def\rf#1{{(\ref{#1})}}

\def\nn{\nonumber }

\def\ket#1{|#1 \rangle}
\def\bra#1{\langle #1|}

\def\2q{{{\{}2{\}}_q}}
\def\3q{{{\{}3{\}}_q}}

\def\pslash{/\!\!\!{p}}

\def\Qslash{/\!\!\!{Q}}

\def\go{\rightarrow  }
\def\endauthors{}
\def\authors#1\endauthors{#1}
\begin{document}

\baselineskip=21pt




\begin{center}
{\Large \bf Double beta decay of $\Sigma^-$ hyperons}\\
\vspace{1.3cm}

{\large C. Barbero$^1$, G. L\'opez Castro$^2$ and A. Mariano$^1$}\\

{\it $^1$ Departamento de F\'\i sica, Facultad de Ciencias Exactas, \\
Universidad Nacional de la Plata, cc 67, 1900 La Plata, Argentina} \\
{\it $^2$ Departamento de F\'{\i}sica, Cinvestav del IPN \\
Apartado Postal 14-740, 07000 M\'exico D.F. M\'exico}
\end{center}
\vspace{1.3cm}

\begin{abstract}
We compute the strangeness-conserving double beta decay
of $\Sigma^-$ hyperons, which is the only hadronic system that can
undergo such decays. We consider both, the lepton number conserving
$\Sigma^-\go \Sigma^+e^-e^-\bar{\nu}\bar{\nu}$
($\beta\beta_{\Sigma_{2\nu}}$) and the lepton number violating
$\Sigma^-\go \Sigma^+e^-e^-$ ($\beta\beta_{\Sigma_{0\nu}}$) modes.
The branching ratios of these $\beta\beta$ decays are suppressed
at the level of $10^{-30}$ considering a light neutrino scenario
in the case of the $\beta\beta_{\Sigma_{0\nu}}$ channel. The
dynamical origin of such low rates and their possible enhancements
are briefly discussed. Given its simplicity those decays can be
used also for the purposes of illustrating the main features of
double beta decays.
\end{abstract}



\baselineskip=18pt
\bigskip
\noindent
\vspace{0.5in}

\baselineskip=21pt

\begin{center}
\large \bf 1. Introduction
\end{center}

\

  Neutrinoless double beta ($\beta\beta_{0\nu}$) decays would occur
if a mechanism allows the violation of the total lepton number $L$ by two
units. Their observation in experiments will provide unambiguous evidence
for physics beyond the Standard Model (SM). At  present, the first
evidences for physics beyond the SM come from the flavor oscillations in
the neutrino sector that required to explain the deficit observed in solar
\cite{sno} and atmospheric neutrinos \cite{atmospheric}. Flavor
oscillations of neutrinos do not require a change of the
total lepton number (namely $|\Delta L|=0$), allow
us to conclude that neutrinos are massive, but do not establish whether
they are Dirac or Majorana particles. If neutrinos are  Majorana
particles, $\beta\beta_{0\nu}$ decays have to occur at some level; their
observation will establish the Majorana nature of neutrinos beyond any
doubt.

   The strangeness-conserving ($\Delta S=0$) $\beta\beta_{0\nu}$ decays
have been unsuccessfully searched in nuclear transitions for
several decades. An intense activity in the experimental and
theoretical fronts\cite{Hax84,Doi85,Ver86,Tom86,Doi93,Bar98}
witness the importance of such decays as a sensitive probe of
physics beyond the SM.  Some examples of extensions of the SM that
can induce contributions to $\beta\beta_{0\nu}$ decays are
right-handed weak couplings as well as the Higgs exchange
\cite{Moh81}, right-handed weak coupling involving heavy Majorana
neutrinos \cite{Hal76}, massless Majoron emission
\cite{Doi85,Gel81,Bur93,Bam95,Bar96}, and R-parity breaking in the
supersymmetric models \cite{Ver87,Hir95}. From a theoretical point
of view, $\beta\beta_{0\nu}$ decays in nuclei are limited in
precision due a wide range of model-dependent predictions for the
nuclear wavefunctions. In the present work, we study the double
beta (both the  lepton number-conserving $\beta\beta_{2 \nu}$ and
lepton number-violating $\beta\beta_{0\nu}$)
strangeness-conserving decays of the $\Sigma^-$ hyperon. The
$\Sigma^-$ hyperon is a unique system in hadron physics that can
undergo strangeness-conserving double beta decays as it will be
explained below.  The hadronic matrix elements necessary for such
calculations are well known and, therefore, can exhibit the
underlying mechanisms for double beta decays in a more clean way.

  As we have mentioned before, the determination of an upper bound for the
effective neutrino mass in nuclear $\beta\beta_{0\nu}$ decays is limited
by model-dependent evaluations of the nuclear matrix elements. Some of the
difficulties we encounter in those calculations are the following:
\begin{enumerate}
\item{}The nucleus  is a many body system with many degrees of freedom;
in practice there is not a well defined rule to choose the most relevant
components to describe an specific excitation;

 \item{}The Hilbert space where nuclear models are worked out have a huge
dimension requiring a lot of time consuming computational work; and

\item{}The multipole expansion for the $\beta\beta_{0\nu}$ decay amplitude
is rather complex making theoretical expressions difficult to manipulate.
\end{enumerate}

  Despite these limitations in theoretical inputs, the large sensitivity
of present experiments have been able to set strong constrains on the
so-called effective Majorana mass term, $\langle m_{ee} \rangle \equiv
\sum U_{el}^2m_{\nu_l}$, where $m_{\nu_l}$ denote neutrino mass eigenstates.
By assuming that $\beta\beta_{0\nu}$ in nuclei are
mediated by the exchange of light Majorana neutrinos, the experimental
upper bound on $\langle m_{ee} \rangle$ is $0.2$ eV \cite{Bau99}.

  Before we proceed with our calculation, it is interesting to take a look
at other reactions that can provide information on the violation of lepton
number. The upper limits available on these rare processes can be used to
set upper limits on the matrix elements $\langle m_{\alpha \beta} \rangle
\equiv \sum U_{\alpha l}U_{\beta l}m_{\nu_l}$, where $\alpha,\beta = e, \mu,
\tau$.  Thus,  muon to positron conversion in nuclei
$\mu^-+(A,Z)\rightarrow (A,Z-2)e^+$ \cite{Doi85,Kau98} gives
$\langle m_{e\mu}\rangle <17(82)$ GeV depending on the spin of the
initial proton pair; the production of three muons in neutrino-nucleon
scattering \cite{Fla00a} leads to the upper limit  $\langle
m_{\mu\mu}\rangle \lsim 10^4$ GeV; this limit has been slightly improved
at HERA  through the reaction $e^+p\go \bar{\nu}l_1^+l_2^+ X$
\cite{Fla00b}, giving  $\langle m_{\mu\mu}\rangle \lsim 4\times 10^3$ GeV
and also for first time limits on   the $\langle m_{l\tau}\rangle$
(connected with the $\tau$-sector)
were given; finally, the non-observation of heavy Majorana neutrinos
at various colliders \cite{Cve99} can also be used to set limits on the
effective Majorana mass. Similarly, some rare kaon decays are also
useful to constrain lepton number violating interactions. For
instance, present bounds on the branching ratio of the
$K^+\go\mu^+\mu^+\pi^-$ decay \cite{Zub00} translates into the upper
limit $ \langle m_{\mu\mu} \rangle \lsim 4\times 10^4$ MeV \cite{Lit00}.
Those bounds from collision experiments and rare kaon decays are
several orders of magnitude above  the limit $ \langle m_{\mu\mu}
\rangle \leq 4.4$ eV,
inferred by an analysis \cite{Barg98} that combines
experimental constrains  from atmospheric and solar neutrino
oscillation and the tritium beta decay end-point experiment \cite{Wei99,Lob99}.

Conversely, we can use the existing bounds from lepton number violating
processes to set upper limits on the branching ratios of some rare kaon
decay. Thus, the upper limit $B(K^+\go\mu^+\mu^+\pi^-)\lsim
10^{-30}\ (10^{-19})$ \cite{Dib00} can be obtained in models with a light
(heavy) neutrino scenario, while  $B(K^+\go e^+ e^+\pi^-)\lsim
10^{-26}$ \cite{Lit92} can be derived from upper bounds from nuclear
$\beta \beta_{0\nu}$ decays. These values are well below the
sensitivities of current experiments; for example the limits reported by
the E865 experiment \cite{Lit00} are  $B(K^+\go e^+ e^+\pi^-)< 6.4 \times
10^{-10}$, and $B(K^+\go \mu^+
\mu^+\pi^-)\lsim 3\times 10^{-9}$. This shows that we are far from
detecting such  processes, in spite of the special window in the  hundred
MeV region \cite{Dib00},  and that the sensitivity of such processes are well
below the previously discussed $\beta\beta_{0\nu}$ case.  Nevertheless it
is important to pursue searches for $|\Delta L|=2$ processes since they
would lead to nonvanishing results even if nuclear $\beta\beta_{0\nu}$
decay turns out to vanish or becomes extremely suppressed.

In this Letter we study the two double beta decays of the
$\Sigma^-$ hyperon, namely: the lepton number conserving
$\Sigma^-\go \Sigma^+e^-e^-\bar{\nu}\bar{\nu}$
($\beta\beta_{\Sigma_{2\nu}}$) and the lepton number
violating $\Sigma^-\go \Sigma^+e^-e^-$
($\beta\beta_{\Sigma_{0\nu}},\ \Delta L=2$) channels. The Feynman graphs
for the $\beta\beta_{\Sigma_{2\nu}}$ and
$\beta\beta_{\Sigma_{0\nu}}$ decays, indicating the two intermediate
states, are shown in Figure 1a and 1b, respectively. To our
knowledge, these decay modes have not been reported previously
in the literature and even an experimental upper limit is not available
(an upper bound of order $10^{-4}$ was reported in ref. \cite{Lit92b} for the
branching ratio  of the related strangeness-changing $\Sigma^- \rightarrow p
\mu^-\mu^-$ double beta decay).
The isotriplet of $\Sigma$ hyperons ($\Sigma^+,\ \Sigma^0,\ \Sigma^-$) is
a unique system of hadrons that can undergo strangeness-conserving double
beta decays. Actually,
the $\Sigma^-$ and the $\Sigma^+$ are not antiparticles of each other;
they have a mass splitting of $m_{\Sigma^-}-m_{\Sigma^+}=8.08$ MeV \cite{PDG00},
which allows a sufficient phase space for $\beta\beta$ decays. An
interesting feature of this system is that we can identify {\it only two}
well defined intermediate states ($\Sigma^0$ and $\Lambda$) that can give
important contributions to the $\beta\beta$ transitions under
consideration. One of these states (the $\Sigma^0$) lies in the middle of
the $\Sigma^-,\ \Sigma^+$ levels while the other (the $\Lambda$) is around
77 MeV below them. Namely, one of these intermediate states can be real
while the other is virtual.

  On another hand, the hadronic matrix elements required for the
calculations are dominated by a couple of axial and vector form factors,
which have been computed by several authors in the literature
\cite{Pio85,Sch95}
with good agreement among them. This is an important advantage over the
evaluation of nuclear transition matrix elements. All these
characteristics of the $\Sigma$ hyperon system make very interesting the
study of their double beta decays, even as a possible textbook example to
learn the formalism of $\beta \beta$ transitions applied to free baryons.
In the following we consider each case separately.

\

\begin{center}
{\large \bf 2. Lepton number-conserving $\beta\beta$ decay}
 \end{center}

   One of the four Feynman graphs corresponding to the $\beta\beta_{\Sigma
2\nu}$ decay is shown in Fig. 1a (the other three contributions are
obtained under proper antisymmetrization with respect to final state
electrons and antineutrinos).

\begin{figure}[h]
\vspace{-0.5cm}
\begin{center}
    \leavevmode
   \epsfxsize = 14cm
     \epsfysize = 9cm
    \epsffile{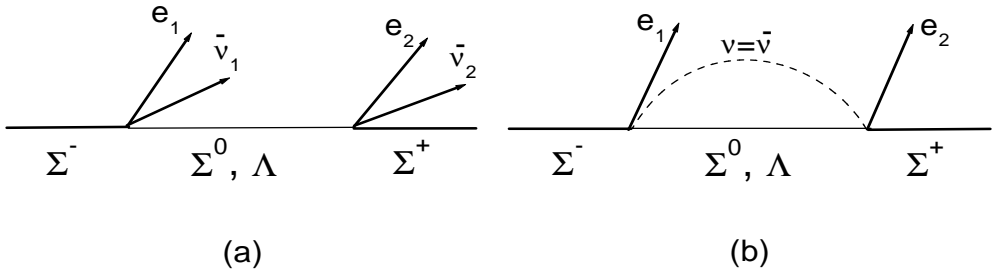}
   \end{center}
\vspace{-3cm}
\caption{Lowest order diagrams contributing to the $\Sigma^-$ decay for
(a) two neutrino double beta decay mode, and (b) neutrinoless double beta
decay mode.}
\label{fig1}
\end{figure}

The effective four-fermion weak Hamiltonian acting at each vertex has
the usual current-current form \cite{Bar98,Sch95}
\be
H_W=\frac{G}{\sqrt{2}}J_\mu j^\mu+h.c.
\label{1}\ee
where
\be
J_\m=\bar{\psi}_{B'}\g_\m(f_1+g_1\g_5)\psi_{B},
\label{2}\ee
 is the baryonic current operator underlying the $B\rightarrow
B'$ transition and
\be
j_\m=\bar{\psi}_e\g_\m(1-\g_5)\psi_\n, \label{3}
\ee
is the $V-A$ leptonic current operator\footnote{We do not
consider here the possible existence of  right-handed neutrinos, since
their  contribution to the nuclear double beta decay is negligible
\cite{Doi85,Bar98}.}. In eq. (2), $f_1$ and $g_1$ are dimensionless
vector and axial-vector form factors for the $B\rightarrow B'$
transition. In our approximation we have neglected their momentum
transfer dependence and we have also neglected the small
contributions of the induced magnetic and scalar form factors for the
vector and axial currents. The effective weak coupling constant is
$G=G_F V_{ud}$, where  $G_F=(1.16639\pm 0.00003)\x 10^{-11}$ MeV$^{-2}$
is the Fermi constant and $V_{ud}$ the relevant element of the
Cabibbo-Kobayashi-Maskawa mixing matrix.

The rate  for the $\beta\beta_{\Sigma_{2\nu}}$ decay from the
initial $\Sigma^-\equiv A$ to the final $\Sigma^+\equiv B$ hyperon is
given in the rest frame of $A$ by (we use natural units, \ie
$\hbar=c=m_e=1$)
\be
d\Gamma_{\Sigma 2\n}=\pi
\sum_{spin}|{\cal M}_{2\n}|^2
\delta(m_A-\e_B-\e_1-\e_2-\w_1-\w_2)\frac{d\pb_1}{(2\pi)^3}
\frac{d\pb_2}{(2\pi)^3}\frac{d\qb_1}{(2\pi)^3}\frac{d\qb_2}{(2\pi)^3},
\label{4}\ee
where $m_A$ ($\e_B$) is the energy of the initial (final) hyperon,
and $\e_i$ and $\pb_i$ ($\w_i$ and $\qb_i$) denote the energy and momentum
of the electron (neutrino). The decay amplitude reads

\br {\cal M}_{2\n}&=&[1-P(e_1e_2)][1-P(\n_1\n_2)]\nonumber \\ & &
\sum_{\eta=\Sigma^0,
\Lambda}\frac{\bra{p_B;e_1e_2\tilde{\nu}_1\tilde{\nu}_2}H_W
\ket{\eta,Q;e_1\tilde{\nu}_1}\bra{\eta,Q;e_1\tilde{\nu}_1}H_W\ket{p_A}}
{m_A-\e_{Q}(\eta)-\e_1-\w_1}, \label{5}\er
with $\e_{Q}(\eta)=\sqrt{\Qb^2+m_\eta}$ and $\Qb\equiv
-\pb_1-\qb_1=\pb_B+\pb_2+\qb_2$ being the energy and
momentum of the intermediate baryon state, and $P(x_1x_2)$ the
operator that exchanges $x_1$ with $x_2$. Thus, after introducing
eqs. \rf{1}--\rf{3} in eq. \rf{5} we  get

\be
{\cal M}_{2\n}=\frac{G^2}{2}[1-P(e_1e_2)][1-P(\n_1\n_2)]\sum_{\eta=\Sigma^0,
\Lambda}\frac{
B_\m(p_A,Q;A)B_\n(Q,p_B;B)L^{\m\n}(p_1,q_1,p_2,q_2)}
{m_A-\e_{Q}(\eta)-\e_1-\w_1},
\label{6}\ee
where
\be
B_\m(p_I,p_F;X)=\bar{u}(p_F)\g_\m(f_{X\eta}+g_{X\eta}\g_5)u(p_I),
\label{7}\ee
is the baryonic matrix element between states with momentum
$p_I$ and $p_F$, and
\be
L_{\m\n}(p_1,q_1,p_2,q_2)=\bar{u}_e(p_2)\g_\m(1-\g_5)u_{\tilde{\n}}(q_2)
\bar{u}_e(p_1)\g_\m(1-\g_5)u_{\tilde{\n}}(q_1), \label{8}\ee is
the leptonic tensor. The values for the form factors
$f_{A\eta} \equiv f_1({\Sigma^-\eta})$,
$g_{A\eta} \equiv g_1({\Sigma^-\eta})$, $f_{B\eta} \equiv
f_1({\eta\Sigma^+})$ and $g_{B\eta} \equiv g_1({\eta\Sigma^+})$ at zero
momentum transfer are summarized in Table \ref{tab1}:

\begin{table}[h]
\begin{center}
\caption {Hyperon form factors at zero momentum transfer (see Table III on
Ref. \cite{Sch95}).}
\label{tab1}
\bigskip
\begin{tabular}{c|cccc}
\hline
\hline
$\eta$&$f_{A\eta}$&$g_{A\eta}$&$f_{B\eta}$&$g_{B\eta}$\\
\hline
$\Lambda$&$0$&$-0.60$&$0$&$-0.60$\\
$\Sigma^0$&$1.41$&$-0.69$&$-1.41$&$0.69$\\
\hline\hline \end{tabular} \end{center}
\end{table}

Given the small mass difference between the relevant hyperon states,
we can use the non-relativistic impulse approximation for the baryonic
current \cite{Bar98,Bar95}. The small mass difference between the $\Sigma$
hyperon states is responsible of the suppression of the decay via the real
intermediate $\Sigma^0$ particle, while the  decay through an
intermediate on-shell $\Lambda$ state is forbidden since $m_\Lambda <
m_\Sigma^+$. Keeping only the usually called Fermi and Gamow-Teller
operators we have
\be
B_\m(p_I,p_F;X)= \chi_{m_{s_F}}^\d(f_{X\eta}g_{\m 0}-g_{X\eta}\sigma_k
g_{\m k}) \chi_{m_{s_I}},
\label{9}\ee
where $s_I$ and $s_F$ denote the spin of the initial and final baryons in
the $I\rightarrow F$ transition. In the spirit of the non-relativistic
approximation the energy denominator in eq.
\rf{6} can be also simplified making $\e_{Q}(\eta)\cong m_\eta$.
Note that in the case of the $\Sigma^0$ intermediate state, eq. (6)
exhibits a singularity when $\epsilon_1+\omega_1=m_A-m_{\Sigma^0} = 4.8$
MeV (note that this singularity does not appear for the $\Lambda$
intermediate state because $m_A-m_{\Lambda}=81.7\ \mbox{\rm MeV} \gg
\epsilon_1+\omega_1$). This singularity can be cured by taking into
account the finite width ($\Gamma_{\Sigma^0}=8.89$ KeV \cite{PDG00}) of
the $\Sigma^0$
intermediate state. Therefore, we will define:
\begin{itemize}
\item{}In the case $\eta=\Sigma^0$:
\br
\frac{1}{m_A-m_{\Sigma^0}-\e_1-\w_1}& \rightarrow &
\frac{1}{m_A-m_{\Sigma^0}-\e_1-\w_1+i\frac{\Gamma_{\Sigma^0}}{2}}
\equiv h_{\Sigma^0}(\e_1,\w_1).
\label{10}\er

\item{}In the case $\eta=\Lambda$:
\br
\frac{1}{m_A-m_{\Lambda}-\e_1-\w_1}&\equiv&h_\Lambda(\e_1,\w_1).
\label{11}\er
\end{itemize}

After a lengthy but straightforward calculation we can write the decay
rate as follows:
\be
\Gamma_{\Sigma 2\n}=\frac{G^4}{8\pi^{7}}
\sum_{j=1}^3\sum_{\eta\eta'}\C_j(\eta\eta')\I_j(\eta\eta')\ .
\label{12}\ee
The $\C_j(\eta\eta')$ terms denote quartic combinations of form factors,
\br
\C_1(\eta\eta')&=&f_{A\eta}f_{B\eta}f_{A\eta'}f_{B\eta'}
+3g_{A\eta}f_{B\eta}g_{A\eta'}f_{B\eta'}
+3f_{A\eta}g_{B\eta}f_{A\eta'}g_{B\eta'}
+9\frac{}{}g_{A\eta}g_{B\eta}g_{A\eta'}g_{B\eta'},\nn\\
\C_2(\eta\eta')&=&
f_{A\eta}f_{B\eta}f_{A\eta'}f_{B\eta'}
+3g_{A\eta}f_{B\eta}f_{A\eta'}g_{B\eta'}
+3f_{A\eta}g_{B\eta}g_{A\eta'}f_{B\eta'}
-3g_{A\eta}g_{B\eta}g_{A\eta'}g_{B\eta'},\nn\\
\C_3(\eta\eta')&=&
f_{A\eta}f_{B\eta}f_{A\eta'}f_{B\eta'}
+3f_{A\eta}f_{B\eta}g_{A\eta'}g_{B\eta'}
+3f_{A\eta'}f_{B\eta'}g_{A\eta}g_{B\eta}
+3g_{A\eta}f_{B\eta}f_{A\eta'}g_{B\eta'}\nn\\
&+&3g_{A\eta'}f_{B\eta'}f_{A\eta}g_{B\eta}
+3g_{A\eta}f_{B\eta}g_{A\eta'}f_{B\eta'}
+3f_{A\eta}g_{B\eta}f_{A\eta'}g_{B\eta'}
-3g_{A\eta}g_{B\eta}g_{A\eta'}g_{B\eta'},\nn\\
\label{13}\er
and $\I_j(\eta\eta')$ are the phase space factors defined as follows:
\br
\I_1(\eta\eta')&=&\int_{1}^{\e_0-1} p_1^2 dp_1
\int_{1}^{\e_0-\e_1} p_2^2 dp_2
\int_{0}^{\e_0-\e_1-\e_2}q_1^2  (\e_0-\e_1-\e_2-q_1)^2dq_1\nn\\
&\x&h_{\eta}(\e_2,\e_0-\e_1-\e_2-q_1)h_{\eta'}^*(\e_2,\e_0-\e_1-\e_2-q_1),
\nn\\
\I_2(\eta\eta')&=&\int_{1}^{\e_0-1} p_1^2 dp_1
\int_{1}^{\e_0-\e_1} p_2^2 dp_2
\int_{0}^{\e_0-\e_1-\e_2}q_1^2  (\e_0-\e_1-\e_2-q_1)^2dq_1\nn\\
&\x&Re\left[h_{\eta}(\e_2,\e_0-\e_1-\e_2-q_1)h^*_{\eta'}(\e_1,q_1)\right],\nn\\
\I_3(\eta\eta')&=&\int_{1}^{\e_0-1} p_1^2 dp_1
\int_{1}^{\e_0-\e_1} p_2^2 dp_2
\int_{0}^{\e_0-\e_1-\e_2}q_1^2  (\e_0-\e_1-\e_2-q_1)^2dq_1\nn\\
&\x&Re\left[h_{\eta}(\e_2,\e_0-\e_1-\e_2-q_1)h^*_{\eta'}(\e_1,\e_0-\e_1-\e_2-q_1)
\right],
\label{14}\er
where $\e_0\equiv m_A-m_B$.

  The numerical values for the factors entering in the expression of the
decay rate, eq. (12), are given in Table 2. We can check that the main
contribution comes from the term ${\cal C}_1(\Sigma^0\Sigma^0){\cal
I}_1(\Sigma^0\Sigma^0)$, which includes the contribution of a real
$\Sigma^0$ hyperon intermediate state.

\begin{table}[h]
\begin{center}
\caption {Numerical values for $\C_j(\eta\eta')$ and $\I_j(\eta\eta')$ (in
units of MeV$^9$) for $\b\b_{\Sigma_{2\nu}}$ decay.}
\label{tab2}
\bigskip
\begin{tabular}{c|cc|cc|cc}
\hline
\hline
$j$&$\C_j(\Lambda\Lambda)$&$\I_j(\Lambda\Lambda)$&
$\C_j(\Sigma^0\Sigma^0)$&$\I_j(\Sigma^0\Sigma^0)$&
$\C_j(\Lambda\Sigma^0)=\C_j(\Sigma^0\Lambda)$&$\I_j(\Lambda\Sigma^0)+\I_j(\Sigma^0\Lambda)$\\
\hline
$1$&$1.166$&$5.46\x 10^{-1}$&$11.672$&$6.740\x 10^5$&$-1.543$&$4.106\x 10$\\
$2$&$-0.389$&$5.46\x 10^{-1}$&$8.952$&$2.162 \x 10^3$&$0.514$&$3.252 \x 10^2$\\
$3$&$-0.389$&$5.46\x 10^{-1}$&$20.3101$&$4.598 \x 10^3$&$-1.633$&$4.170\x 10$\\
\hline\hline \end{tabular} \end{center}
\end{table}
   Using the values obtained in Table 2, we can compute the branching
ratio from the rate in eq. (12). We obtain:
\begin{equation}
B(\beta\beta_{\Sigma_{2\nu}})= 1.38 \times 10^{-30}\ (1.36 \times
10^{-30})\ .
\end{equation}
Just for comparison we have shown within parenthesis the value
corresponding to the contribution of the $\Sigma^0$ intermediate state. As
expected, this contribution dominates almost completely the decay rate.
The branching ratio given above turns out to be very suppressed due
essentially to the large decay width of the $\Sigma^0$ hyperon appearing
in eq. (10). As we know, the decay rate of the $\Sigma^0$ is 10 orders of
magnitude larger than those of the charged $\Sigma$ hyperons because it
can undergo the electromagnetic decay $\Sigma^0 \rightarrow \Lambda
\gamma$.

\

\begin{center}
\large \bf 3. Neutrinoless double beta decay
\end{center}

The decay rate for the (three-body) neutrinoless
$\beta\beta_{\Sigma_{0\nu}}$ mode reads
\be
 d\Gamma_{\Sigma 0\n}=\pi\sum_{spin}|{\cal
M}_{0\n}|^2\delta(m_A-\e_B-\e_1-\e_2)
\frac{d\pb_1}{(2\pi)^3}\frac{d\pb_2}{(2\pi)^3}\ .
\label{15}\ee
The decay amplitude corresponding to the diagram in Figure 1b (after
proper anti-symmetrization with respect to identical electrons) is
\br
{\cal M}_{0\n}&=&[1-P(e_1e_2)]\sum_{\eta=\Sigma^0,
\Lambda}\sum_{s_\nu}\int\frac{d^4q}{(2\pi)^4}
\nonumber \\
&&\frac{\bra{p_B;e_1e_2}H_W\ket{\eta,Q(q);e_1\tilde{\nu}}
\bra{\eta,Q(q);e_1\tilde{\nu}}H_W\ket{p_A}}
{m_A-\e_1-q-\e_{Q}(\eta)}\ .
\label{16}\er
The four-momentum of the intermediate state $\eta$ is $Q(q)\equiv
p_A-p_1-q=p_B+p_2-q$ and its energy is
$\e_{Q}(\eta)=\sqrt{\Qb^2(\qb)+m_\eta^2}$. Introducing \rf{1} into
\rf{16}, and expanding the weak neutrino eigenstate as a mixture
of light massive Majorana neutrino states, \ie
$u_{\tilde{\n}}(q)=\sum_{l}U_{el}u_{\n_l}(q)$, we get
\br
{\cal M}_{0\n}&=&G^2[1-P(e_1e_2)]\label{17}\\
&\x&\sum_{l}m_{\nu_l}U_{el}^2
\sum_\eta
\bar{u}(p_B)\g_\m(f_{B\eta}+g_{B\eta}\g_5)I_\eta(p_1)
\g_\n(f_{A\eta}+g_{A\eta}\g_5)u(p_A)L^{\m\n}(p_1,p_2),\nn
\er
with the leptonic factor
\br
L_{\m\n}(p_1,p_2)&=&\bar{u}_e(p_1)\g_\m(1-\g_5)\g_\n u_e^C(p_2)\ .
\label{18}\er
The factor $I_\eta(p_1)$ in eq. \rf{17} corresponds to the following loop
integral ($Q(q)=p_A-p_1-q$):
\be
I_\eta(p_1)=\int \frac{d^4q}{(2\pi)^4}
\frac{\Qslash(q)+m_\eta}{(q^2-m_{\nu_l}^2)(Q^2(q)-m_\eta^2)},
\label{19}\ee which has a logarithmic divergence that we will
manipulate in a simple cutoff procedure. After using Feynman
parametrization techniques  \cite{Gro}, we get \be
I_\eta(p_1)=\frac{i}{8\pi^2}\int_0^1 dx
[(\pslash_A-\pslash_1)x+m_\eta]\int_0^{\Lambda_c}
\frac{k^3dk}{(k^2+M^2)^2}, \label{20}\ee where $\Lambda_c$ is the
cutoff energy and we have defined
\be
M^2 = m_\eta^2 (1-x)-(p_A-p_1)^2(1-x)x+m_{\n_l}^2x\ .
\label{21}\ee
We stress here that the origin of this logarithmic divergence
is related to the effective vertices we are using for the hadronic
form factors. This divergence can in principle be cured by including the
weak form factors which are expected to fall as $1/[(p_A-Q(q)]^2$ with
$q^2 \rightarrow \infty$ in the dipole approximation, but their real
behavior in
the large $q^2$ limit are actually determined by QCD.
In the case of the nuclear $\beta\beta_{0\nu}$ decays it is usual to
assume that the divergence is driven by the internucleon distance, which
sets $\Lambda_c \sim 1$ GeV. This value can be identified with the
largest momenta that the neutrino can carry which in turn is fixed
by the lowest distance between two nucleons in the nucleus ($\Lambda_c
\sim (2d)^{-1}$, where $d$ is the nucleon radius). We will assume here
that  the average distance between the quarks within the hyperons are of
the order of a typical hyperon (or nucleon) radius. Here, we adopt the
point of view that our results are stable as far as they do not depend
strongly on the specific value of the cutoff $\Lambda_c$.

The integration in eq. \rf{20} can be  simplified by neglecting in
eq. \rf{21} all lepton masses and momenta, which is consistent for the
neutrino since we are assuming a light neutrino scenario. We obtain
\be
I_\eta(p_1)=\frac{i}{(4\pi)^2}
\left[\frac{}{}(\pslash_A-\pslash_1)F_\eta+m_\eta G_\eta\right],
\label{22}\ee
where
\br
F_\eta&=&\frac{1}{4}(1+m^2)\mbox{ln}(m+m')-\fot(1- m^2)\mbox{ln}(1-m)
-\fot m^2\mbox{ln}(m)+\frac{1}{4}(1-m^2)\mbox{ln}(m')\nn\\
&-&i\frac{\pi}{2}(1+m^2)+\frac{[2m'-(1-m)^2](1+m)}{2D}\left[
arctg\left(\frac{1-m}{D}\right)+arctg\left(\frac{1+m}{D}\right)\right]
,\nn\\
G_\eta&=&\fot(1+m)\mbox{ln}(m+m')-(1- m)\mbox{ln}(1-m)-m\mbox{ln}(m)+\fot(1-m)\mbox{ln}(m')\nn\\
&-&i\pi(1+m)+\frac{2m'-(1-m)^2}{D}\left[arctg\left(\frac{1-m}{D}\right)+
arctg\left(\frac{1+m}{D}\right)\right]\ .
\label{23}\er
In the above expressions, we have introduced the following
dimensionless constants:  $m={m_\eta^2}/{m_A^2}$,
$m'={\Lambda_c^2}/{m_A^2}$ and $D=\sqrt{4m'-(1-m)^2}$.

In order to evaluate the unpolarized squared amplitude we will use
the non-relativistic impulse approximation for the final baryon
and  take $p_B\simeq(m_B,\bf{0})$ (we can not do the same
approximation for the intermediate baryon state in this case). The
decay rate for the $\beta\beta_{0\nu}$ transition becomes: \br
\Gamma_{\Sigma 0\n}&=&\langle m_{ee}\rangle ^2\frac{G^4}{4\pi^7}
\I_{0\n}\nonumber\\
&\x&\sum_{\eta\eta'}\left[m_A^2F_{\eta}F_{\eta'}^*\D_1({\eta\eta'})
-2m_Am_\eta (G_{\eta}F_{\eta'}^*+G_{\eta}^*F_{\eta'})
\D_2({\eta\eta'}) +4m_\eta
m_{\eta'}G_{\eta}G_{\eta'}^*\D_3({\eta\eta'}) \right],\nn\\
&&\label{24} \er
where $\langle m_{ee} \rangle
=\sum_{l}m_{\nu_l}U_{el}^2$ is the effective neutrino mass . The other
factors appearing in eq. \rf{24} are defined as follows:
\br
\D_1({\eta\eta'})&=&(f_{A\eta}f_{B\eta}+g_{A\eta}g_{B\eta})(f_{A\eta'}f_{B\eta'}+g_{A\eta'}g_{B\eta'})
, \nn\\
\D_2({\eta\eta'})&=&(f_{A\eta}f_{B\eta}-g_{A\eta}g_{B\eta})(f_{A\eta'}f_{B\eta'}+g_{A\eta'}g_{B\eta'})
, \nn\\
\D_3({\eta\eta'})&=&(f_{A\eta}f_{B\eta}-g_{A\eta}g_{B\eta})(f_{A\eta'}f_{B\eta'}-g_{A\eta'}g_{B\eta'}),
\label{25}\er for the product of form factors, and
\be
\I_{0\nu}=\int_{1}^{\e_0-1} \e_1(\e_0-\e_1) \sqrt{(\e_1^2-1)[(\e_0-\e_1)^2-1]} d\e_1.
\label{26}\ee
for the phase space integral.

\begin{figure}[h]
\begin{center}
    \leavevmode
   \epsfxsize = 12cm
     \epsfysize = 10cm
    \epsffile{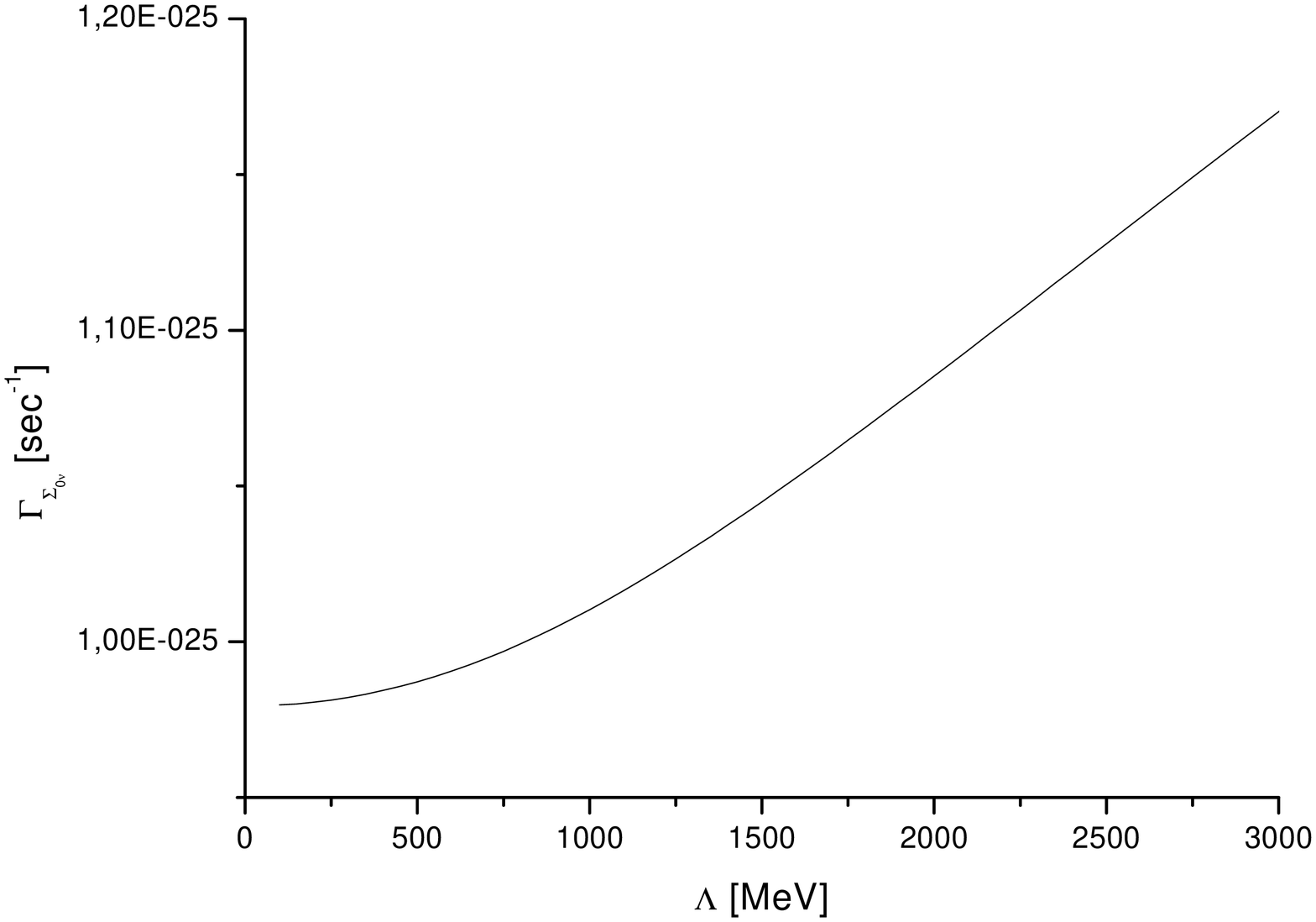}
   \end{center}
\caption{$\beta\beta_{\Sigma 0 \nu}$ decay rate as a function of the
cutoff $\Lambda_c$, for $\langle m_{ee} \rangle=10$ eV.} \label{fig2}
\end{figure}

The neutrinoless double beta decay rate depends on the cutoff
$\Lambda_c$ and the neutrino effective mass $\langle m_{ee}
\rangle$, which are free parameters in our model. Based on present
bounds on electron neutrino mass $\sim eV$ \cite{Bau99}, we show
in Fig. \ref{fig2} the decay rate $\Gamma_{\Sigma_{0\nu}}$ as a
function of the cutoff $\Lambda_c$, for a fixed neutrino mass of
$10$ eV. From this figure we can clearly see that our results are
not very sensitive to the specific cutoff value in the region
under consideration, which gives further support to the  method
employed for the logarithmic divergence.

   For illustrative purposes let us use $\Lambda_c=1$ GeV, which
corresponds to a rough estimate of  the inverse of the  size of
hyperons.  The branching ratio in this case becomes
\begin{equation}
B(\beta\beta_{\Sigma 0\n})= 1.49\times 10^{-35}\ .
\end{equation}
Thus, the neutrinoless double beta decay of the $\Sigma^-$ hyperon is very
suppressed and even smaller that the branching ratios
of $\beta\beta_{0\nu}$ decays of kaons. Note however that in the case of
$\Sigma$ hyperons we do not have a single neutrino
(tree-level) intermediate state contribution as in the case of
neutrinoless double beta kaon decays \cite{Dib00}.

\
\newpage
\begin{center}
\large \bf 4. Summary and conclusions
\end{center}

  We have considered the strangeness-conserving double beta decays of
$\Sigma^-$ hyperons. Several characteristics make this particle a
unique and interesting system to study (and to learn) such decays as an
alternative to the corresponding decays in nuclei. First,
the isotriplet of $\Sigma$ hyperons is the only system of hadrons that can
undergo double beta decays since the $I_3=+1,\ -1$ components have
different masses. Second, we can identify a few intermediate states that
dominate such decays where one state is in the middle and the other is
below the initial and final hadronic levels. Finally, the uncertainties
in hadronic matrix elements are much smaller than in the nuclear case.

  We study both the lepton number-conserving $\beta\beta_{\Sigma 2 \nu}$
and the
lepton number-violating $\beta\beta_{\Sigma 0 \nu}$ decays. For the lepton
number
conserving decay we obtain $B(\beta\beta_{\Sigma 2 \nu}) =1.38 \times
10^{-30}$ and for the neutrinoless decay we get $B(\beta\beta_{\Sigma 0
\nu}) =1.49 \times 10^{-35}$, standing in the light neutrino scenario in
the last
case. The suppression of the lepton number-conserving decay is due to the
large
decay width of the $\Sigma^0$ intermediate state. Therefore, it would be
nice to find an analogue system to the $\Sigma$ hyperons (the $\Sigma_b$
baryons?) where the mechanism under consideration could produce an
enhancement of the decay rate if the corresponding isotriplet is similar
to the spectra of $\Sigma$ hyperons. Our result for the neutrinoless
double beta decay is
almost insensitive to the specific value of the cutoff parameter used to
regulate the divergent integrals.

 Our numerical results for the branching ratios may look discouraging.
However, let us assume an hypothetical model where $B(\beta\beta_{\Sigma
0\nu})=(10^{-20}\sim 10^{-25}) \langle m_{ee} \rangle^2$; in this very
optimistic scenario, an experimental upper limit of $10^{-8}$ would
translate into the interesting bound $\langle m_{ee} \rangle \leq 1\sim
300$ MeV. On another hand, if there exists such a model that causes
$\Sigma^-$ to have faster neutrinoless double beta decays, the SM
background due to the lepton number-conserving double beta decays would
certainly
be very small.

\

{\bf Acknowledgements}.  C.B and A.M. acknowledge the support of ANPCyT
(Argentina) under grant BID 1201/OC-AR (PICT 03-04296), and are fellows of
the CONICET (Argentina). The work of G.L.C. has been partially supported
by Conacyt (M\'exico).


\end{document}